# 3D-printed rotating spinnerets create membranes with a twist

Tobias Luelf[1,2], Deniz Rall[1,2], Tim Femmer[1], Christian Bremer[1], Matthias Wessling[1,2]

[1] *RWTH Aachen University, Chemical Process Engineering, Forckenbeckstrasse 51, 52074 Aachen, Germany*

[2] *DWI - Interactive Materials Research, Forckenbeckstrasse 50, 52074 Aachen, Germany*

**Abstract**

Round hollow fiber membranes are long-established in applications such as gas separation, ultrafiltration and blood dialysis. Yet, it is well known that geometrical topologies can introduce secondary flow patterns counteracting mass transport limitations, stemming from diffusion resistances and fouling. We present a new systematic methodology to fabricate novel membrane architectures. We use the freedom of design by 3D-printing spinnerets, having multiple bore channels of any geometry. First, such spinnerets are stationary to fabricate straight bore channels inside a monolithic membrane. Second, in an even more complex design, a new mechanical system enables rotating the spinneret. Such rotating multibore spinnerets enable (A) the preparation of twisted channels inside a porous monolithic membrane as well as (B) a helical twist of the outside geometry. The spun material systems comprise classical polymer solutions as well as metal-polymer slurries resulting in solid porous metallic monolithic membrane after thermal post-processing. It is known that twisted spiral-type bore channel geometries are potentially superior over straight channels with respect to mass and heat polarization phenomena, however their fabrication was cumbersome in the past. Now, the described methodology enables membrane fabrication to tailor the membrane geometry to the needs of the membrane process.

## 1. Introduction

In membrane filtration processes, round membranes in the shape of hollow fibers as well as flat sheets are well-established. Yet, the challenge remains to improve mass transfer at the membrane surface. In membrane processes with high permeation rates, the retained component accumulates at the membrane surface. If back diffusion or mixing is limited, the retained component builds up and an additional mass transfer resistance emerges and the effective driving potential across the membrane is lowered. Additionally, in case of ultrafiltration, retained components increase the viscosity. This also acts as an additional transport resistance and hence lowers the mass transfer. Finally the overall performance of the process decreases. Measures to counteract these concentration polarization and fouling phenomena are manyfold. One of them being the induction of secondary flows normal to the membrane surface back into the feed channel preventing the establishment of a concentration polarization or fouling layer. These secondary flows emerge when the surface of a membrane causes the flowing feed to absorb a momentum, different then the main flow direction. It is highly desirable to have at hand a freedom of design for membrane extrusion processes through the tailored design of a spinneret. This would then result in the desired membrane geometry. In the scope of our publication we aim to overcome the previously described issues concerning mass transport resistance by varying the geometry of the hollow fiber membrane. This paper describes a methodology to freeform fabricate spinnerets with any channel geometry by additive manufacturing, or 3D printing, combined with a superimposed rotation of the spinneret and the fabrication of polymeric and sintered hollow fiber membranes.

## 2. Background

*2.1. Membrane geometries*

Two membrane geometries are used in general: flat sheet and hollow fiber membranes. To enable membrane operation, the membrane needs to be implemented into a module to define flow conditions. For flat sheet membranes spiral-wound and stagged flat modules are most widely used. Both the feed and permeate channel are equipped with spacers. These spacers define the channel height and additionally act as mixing devices by periodical redirection of the flow field [1, 2]. Investigation and optimization of spacer geometry and orientation in spiral wound modules has been subject to many scientific publications, as reviewed by Schwinge et al. [3–12].

For hollow fiber or extruded orifice membrane geometries, only the membrane itself and its arrangements in a module without any spacer define the flow configuration. The hollow fiber geometry simply has a round cross-section and a linear axial orientation. The small dimension of the fibers and the absence of spacers results in high specific surface areas as compared to flat membrane module types. The mixing of feed flow, showing concentration polarization, and the reduction of this in hollow fiber and tubular membranes is subject of scientific research since almost 20 years [13–20]. Especially, concentration polarization or deposition layers are predominant on the membrane surface in inside-out micro- and ultrafiltration. In contrast to the virgin membrane's hydraulic resistance, better lumen mixing offers significant performance improvement [21]. While in flat sheet membrane modules the flow channels are formed by two different membranes with spacers inserted during module fabrication, the situation is different in hollow fiber membranes. Here, the flow channel is not accessible after membrane formation. Further, the placement of passive mixing elements such as turbulence promotors is possible [22] however not feasible in small size hollow fiber geometries.

For the purpose of increasing surface area, non-circular structured hollow fiber membranes with cross-sectional geometries other then round are described in [23] for outside structures and in [24] for inside structures. Outside structures can also have beneficial transport properties when twisted and used in aerated submerged filtration processes [25]. However, integrated mixing functionalities on the inside are not established well due to difficulties in fabrication. To address this fabrication challenge, recent developments utilize the shape of the membrane itself to induce secondary flow, mixing up the diffusion boundary layer. The flow field is designed to transport retained components back to the bulk flow and thus decrease concentration polarization. A method to fabricate circular curled hollow fibers [26] via the liquid rope coil effect has been presented by Luelf et al., as well as fabrication of hollow fibers with axially sine-shaped but radially circular lumen channels [27]. Such passive mixing in single bore hollow fiber applications has been subject to different studies on mass transfer. Especially, spiral shaped structures are of interest due to their specific flow field. Moulin et al. investigated its potential application in ultrafiltration and found curled orientation of hollow fibers to potentially offer increased fluxes [28]. Furthermore, computational fluid dynamics have been applied to show the mixing potential in spiral shaped membrane channels [15, 16]. Pentair recently developed tubular membranes with increased mixing performance, called X-Flow Helix technology. The X-Flow tubular membranes were evaluated by Wiese et al. using NMR and filtration measurements [29]. Here, a small spiralling ridge or corrugation is directly introduced into the bore channel during membrane fabrication, improving mass transport significantly at turbulent flow conditions.

*2.2. Spinneret design*

Hollow fiber membranes are produced via co-extrusion of a bore fluid and a polymer dope solution through a spinneret. Multilayer co-extruded hollow fibers are also known for different applications [30, 31]. Traditional manufacturing techniques for spinnerets produce round geometries of the main polymer outlet and the bore channel only. As dimensions of the spinnerets are in the range of the produced membrane geometry, sophisticated design of spinnerets would lead to excessive manufacturing costs. If additional heating channels have to be provided, limitations in manufacturing become predominant. In general, manufacturing limitations nowadays dictate the design of spinnerets. It would be highly desirable to become independent of current spinneret manufacturing processes and move towards quick, cheap and flexible design methods enabled by 3D-printing.

Membranes with multiple bore channels are well known for monolithic ceramic membranes. Also hollow fiber membranes with multiple bore channels are available from GE or BASF Inge GmbH or, known as *Multibore®* membranes for ultrafiltration [32]. Researchers also employ this geometry experimenting with different channel amounts in a single fiber [33–37]. There have been attempts to improve the membrane geometries with multiple bore channels with regard to wall thickness distribution [34–36]. All of these studies approach spinneret design from a mainly qualitative perspective. However, manufacturing of these multi-channel spinnerets with traditional manufacturing techniques is complex and expensive. Hence, iterative geometry optimization of spinneret design without rapid prototyping technique is currently not feasible.

Bonyadi and Mackley fabricated flat membranes with multiple bore channels, named micro-capillary film membranes and reported the effect of spinning conditions, such as take up speed and air gap distance on the macroscopic membrane formation [38].

*2.3. Spiralling flow channels*

Spiralling channels are known to enhance mixing and transport limited processes [39]. So-called Dean vortices promote the mass transport [13, 15–18, 21]. In a curved flow arrangement, a moving element is affected both by inertia and centrifugal forces. The latter scales with rotational speed and radius. Thus, an element in the bulk flow is dragged in radial direction of curvature. Elements near the channel wall are slower due to no slip condition at the wall. The interplay of both finally result in a double vortex, also known as dean vortices, enabling additional mixing [15, 16, 39, 40]. The formation of dean vortices is commonly described in a mathematical manner using the Dean number. This dimensionless parameter scales the Reynolds number $Re$ with the radius of curvature.

$$De = Re \cdot \sqrt{\frac{d_i}{D_c}} \quad (1)$$

with

$$D_c = D \left[1 + \left(\frac{b}{\pi D}\right)^2\right] \quad (2)$$

Here $D_c$ represents the diameter of curvature, corrected by the helical pitch $b$. The helical pitch determines the length of a 2 $\pi$ twist. The tube internal diameter is denoted by $d_i$. [41]

Transfer phenomena are often described by dimensionless power-law equations. The correlations for heat and mass transfer are similar due to equal mathematical descriptions of their fundamental principles. There are multiple experimental and theoretical studies on the flow and heat transfer in curved pipes. While the reproduction of all studies is beyond the scope of this paper, Vashisth et al. gave a detailed summary on their findings and the conditions under which they were derived [39].

The flow field was applied to reverse osmosis in a curved membrane duct in an early study by Srinivasan and Tien [42] in 1970. Here, potential reduction of concentration polarization in binary salt solutions was proven by a mathematical approach. Nunge and Adams [43] critically evaluated the studies of Srinivasan and Tien and found concentration polarization reduction that is less intense, but still significant.

Winzeler and Belfort [44] have prooven the existance of secondary vortices in combination with a flat sheet membrane in a spiral channel. They visualized the vortices both optical and by NMR imaging and measured five times increased flux for dairy whey filtration by presence of spiral channels.

Liu et al. [41] investigated the mass transfer in curved hollow fiber membrane with different wind angles (pitches) and compared the results to straight fibers. They found significantly increased mass transfer coefficients up to 3.5 fold. Finally beeing expressed by mass transfer coefficients and Sherwood numbers. Equation 3 represents the Leveque solution for straight tubes.

$$Sh_s = 1.62 \left(\frac{d_i}{L}\right)^{0.33} \cdot Re^{0.33} \cdot Sc^{0.33} \tag{3}$$

Liu et. al found correlations for the Sherwood number as a function of Dean number as expressed in Equation 4. Their findings show dependencies of higher power employing values of $\alpha$ up to 0.55 and $a$ of 2.64 regarding fibers with a wind angle of 45 degree.

$$Sh_c = a \cdot De^{\alpha} \cdot Sc^{0.33} \tag{4}$$

As can be seen by the Dean number exponent, the curved structure offers higher transfer rates as compared to the straight channel alignment. Also Moulin et al. [17] found increased mass, expressed by improvement factors of 2 to 4 in water oxygenation.

In our previous work we developed a manufacturing technique to produce hollow fibers based on rope coil spinning without the need of additional manufacturing steps [26]. While the production itself does not demand changes to existing spinning setups, the fiber geometry intrinsically limits the membrane surface area per module volume in closed module operations.

A methodology for the fabrication of integrally twisted multibores does not exist. Here we present a procedure for rapid iterative spinneret improvement, exemplary shown for tribore spinnerets based on rapid prototyping. The method of 3D-printing spinnerets offers freedom of design in the cross section design of hollow fibers. In the second part of the manuscript we investigate the combination of printed spinnerets and a rotational spinneret movement. Thereby, we create fibers with potential passive mixing properties and no limitation in packing density as compared to conventional hollow fibers. The presented method is evaluated by spinning of polymeric fibers. Based on our preliminary experience on synthesis of porous metallic fibers [45], we also apply the method to produce porous electrodes with previously mentioned properties.

## 3. Experimental

### 3.1. Spinning setup

For hollow fiber production by NIPS (non-solvent induced phase separation), a state-of-the-art hollow fiber spinning line was used. The designed spinnerets were connected to the polymer solution and bore fluid lines via a rotating construction for the production of fibers with twisted bore channels. Figure 1 visualizes the position of the rotating construction in the spinning line. The rotating construction, described in Section 3.3 is located above the coagulation bath, filled with tap water. The fiber is spun through the spinneret, passes the air gap and undergoes phase separation in the coagulation bath. The fiber is held in place around a small guiding wheel and finally taken out of the bath by a pulling wheel. It is placed in additional water baths for solvent exchange for at least 48 hours. Due to the rotation and the newly designed spinneret, two new production parameters are associated as follows:

- Spinneret geometry

- Rotational speed of spinneret

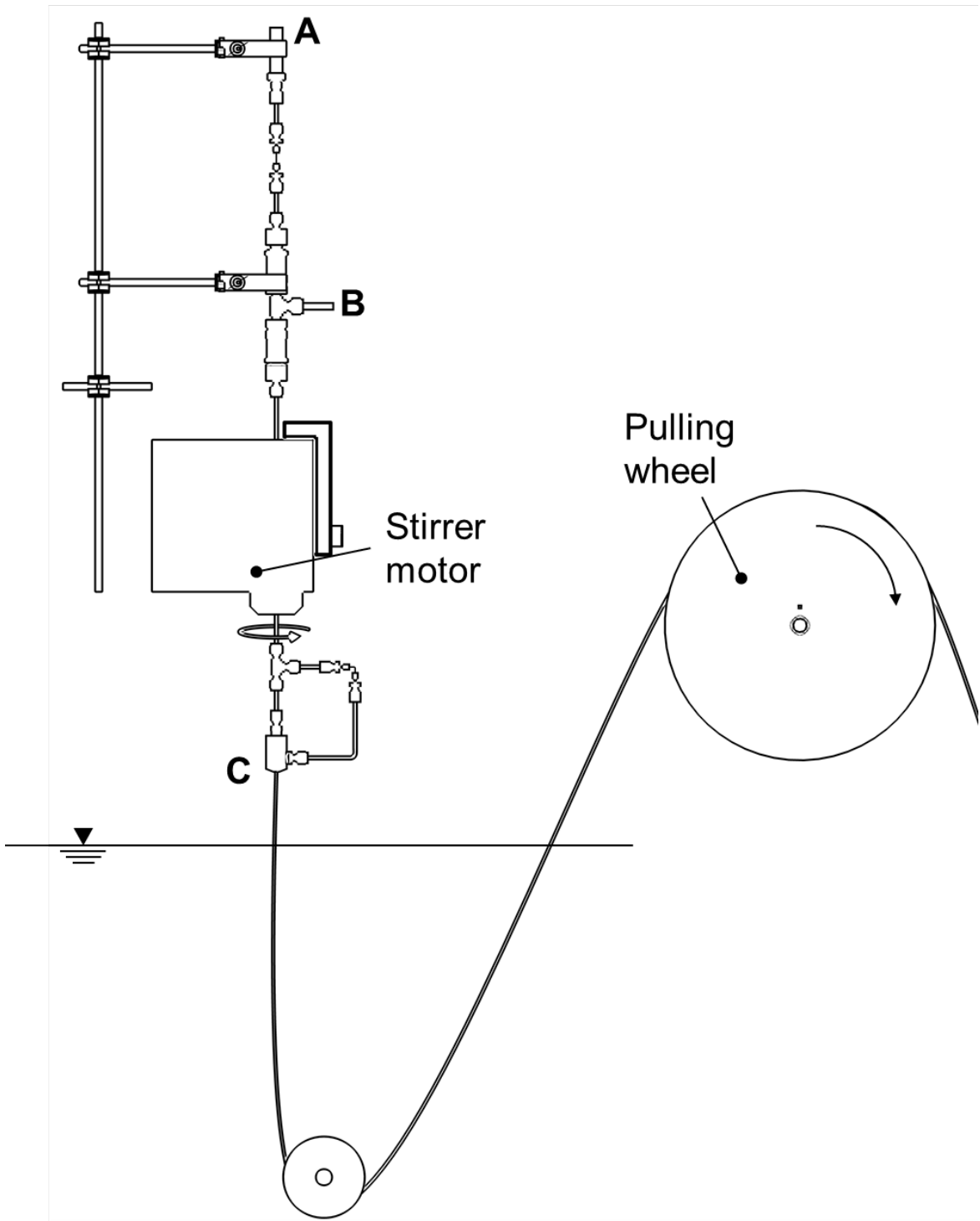

Figure 1. Spinning setup: A) Bore fluid inlet, B) Polymer solution inlet, C) Spinneret outlet.

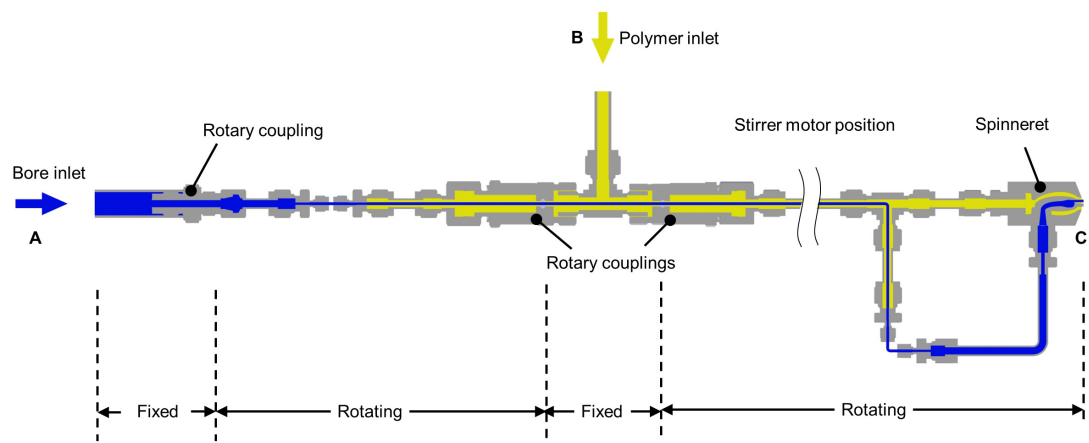

Figure 2. Cut view of rotating spinneret assembly. The bore fluid line is coloured in blue while the polymer solution is guided through the yellow parts. The stirrer motor is located between the cut lines and is not displayed in this view. A) Marks the inlet of bore fluid. B) Marks the inlet of the polymer solution. C) Marks the position of the 3D printed spinneret connected to the bore fluid and polymer solution lines.

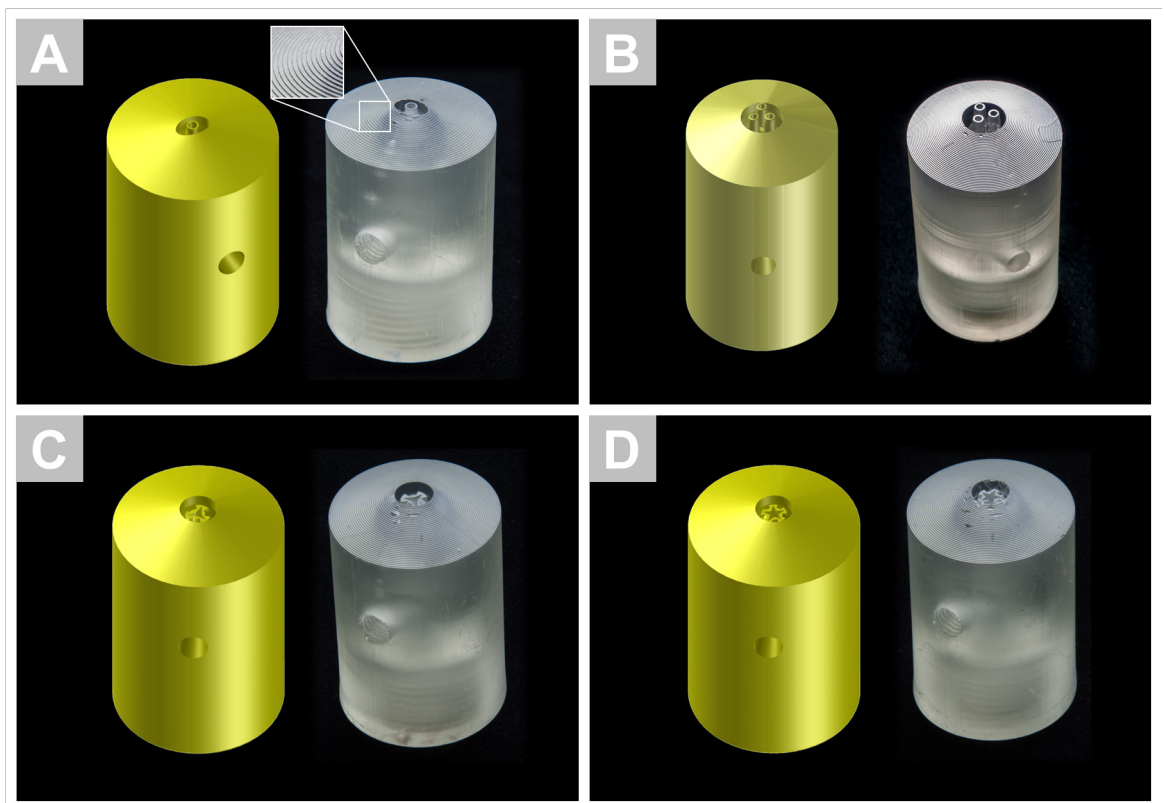

Figure 3. Comparison of spinneret CAD drawings with printed versions. The left images of each pair are CAD renderings, while the right photographs are taken after printing of the spinnerets. Each spinneret body has a diameter of 25mm. A) Oval polymer solution opening with clearly visible step size of 100 $\mu m$ in z-direction shown in the magnification, B) round tri-bore spinneret, C) three edged star shaped bore opening, D) five edged star shaped bore opening.

## 3.2. Rapid prototyping of spinnerets

All spinnerets, except the ones for the production of flat tribore fibers were 3D-printed by means of stereolithography with Envisiontec Perfactory® 3 mini multi lens. EnvisionTEC e-shell 600 was selected as photo resin. The printing resolution in z-direction on was set to 100 $\mu m$ to allow for sufficient accuracy. The x- and y-resolution was set to 60 $\mu m$. The printed slices are cured out of the shallow resin vat with $180 mW/dm^2$ UV emission. The printing process itself operates at a speed of approximately 10mm per hour in z-direction. For post processing of parts printed with e-shell 600 clear EnvisionTEC GmbH states a suitable method [46]. The parts are cleaned in 2-propanol for a few minutes to remove not polymerized material and afterwards dried at 37 ∘C for 30min. Then the part requires to be post cured by the light curing unit Otofash G171 by NK-Optik GmbH with 2000 to 4000 flashes. Figure 3 and 5 show the opening region of a printed spinneret. The step size in z-direction is clearly visible on the top of the spinnerets, while the inner needle is still round and without edges in the horizontal plane.

For the flat tribore fiber production spinnerets were printed with an OBJET Eden260V 3-Dimensional Printing System by Stratasys Ltd., producing parts via the polyjet printing principle. As materials both VeroClear and RGD525 were processed. Closed volume parts that are printed with RGD525 show an infill that is partly filled with support material. Outer shells of the parts are always printed homogeneous from RGD525. For the flat tribore fibers Stratasys VeroClear was used. Employing VeroClear and RGD525, a print can either be performed with support on the entire outer surface of the printed part, leaving a matte surface, or without outer support to result in a glossy surface.

As NMP was used as a solvent for the polymer, all printed parts have been tested for solvent resistance. A detailed study of the swelling behavior and the solvent resistance can be found in the supplementary material.

*3.3. Rotating construction*

In order to enable a rotational movement of the hollow fiber spinneret, a rotation of two fluids is necessary while the supply of these has to be realized at static positions. A rotating system as proposed by Femmer et al. [47] has been used. Figure 2 shows the rotating assembly used in the here presented work. Since the bore fluid has a lower viscosity, it is guided through the inner capillary of the assembly. It is inserted into the setup from the top (Figure 2 A). Rotation starts directly behind the first rotary coupling. The polymer solution is fed from the side opening (Figure 2 B) through the metal tubing. The wider tube diameter accounts for potentially high viscosities of polymer solutions. The polymer solution inlet stands still while the tubing above and below the adjacent rotary couplings are turning. In order to be able to connect standard spinnerets with separated inlets for polymer solution and bore fluids, the bore fluid is guided out of the shell tube directly before entering the spinneret from the side.

### 3.4. Spinning of round rotating PES fibers

A rotating spinneret was assembled into to a spinning line as illustrated in Figure 1. Spinning parameters were set as listed in Table 3. Polymer and bore flow rates have been kept constant as well as the fiber draw speed by the pulling wheel and the air gap. The rotational speed of the spinneret has been varied from 0 RPM, over 30 RPM, to 65 RPM for the tri-bore hollow fibers. The helical tri-bore hollow fibers have been spun at 0 RPM and 30 RPM.

The polymer solution consists of polyethersulfone (PES) (BASF Ultrason 6020 P), NMP (1-Methyl-2-pyrrolidinone, 99%, extra pure, ACROS Organics$^{TM}$), polyvinylpyrrolidone (PVP K90, extra pure, CarlRoth), glycerol (purity $\leq$ 98%, Ph.Eur., anhydrous., Carl Roth). The bore fluid is a mixture of water and glycerol. Polymer solution and bore fluid were used as listed in Table 1.

|  | PES [wt. %] | PVP [wt. %] | NMP [wt. %] | Glycerol [wt. %] | Water [wt. %] |
|---|---|---|---|---|---|
| Polymer | 15 | 5.25 | 75 | 4.75 | - |
| Bore | - | - | - | 93 | 7 |

Table 1. Solution compositions for rotating spinning of polymeric round tri-bore hollow fibers

### 3.5. Spinning of helical fibers

In the next step, both static mixing properties on the outside of the membrane as proposed by Fritzmann et al. [1] and an integrally twisted multibore monolithic membrane are combined. A rotating spinneret assembly described in Section 3.3 is combined with a flat tri-bore spinneret as depicted in Figure 9. In the 3D printed spinneret, the polymer solution or suspension is fed through a tapered hollow space up until exiting the spinneret in the desired dimension of the annular gap. The annular gap of the polymer channel is rounded concentric to the bore channels with spikes between the needles to enable a good distribution of polymer solution around and between the needles. In case of the bore fluid, the total volume flow is divided and fed to three openings of the spinneret, each providing equal volume flow for each needle bore outlet.

Table 2 shows the composition of the spinning solution applied for the production of helical polynmer fibers. The applies spinning parameters are listed in Table 3.

|  | PES [wt. %] | PVP [wt. %] | NMP [wt. %] | Glycerol [wt. %] | Water [wt. %] |
| --- | --- | --- | --- | --- | --- |
| Polymer | 16 | 8 | 76 | - | - |
| Bore | - | - | - | 5 | 95 |

Table 2. Solution compositions for rotating spinning of polymeric helical tri-bore hollow fibers

|  | $V_{Polymer}$ [mL/min] | $V_{Bore}$ [mL/min] | $v_{Pull}$ [mm/s] | $h_{AirGap}$ [mm] |
| --- | --- | --- | --- | --- |
| tri-bore | 5.2 | 2.0 | 29.7 | 15 |
| helical tri-bore | 7.7 | 24 | 3.46 | 5 |

Table 3. Spinning parameters for spinning of polymeric tri-bore and helical tri bore fibers hollow fibers

*3.6. Spinning of rotating titanium fibers*

Analog to the spinning of rotating PES fibers the dope solution for the titanium fibers is polymer based and adapted from the method of our previous work concerning the production of tubular macro-porous titanium membranes by David et al. [45].

*3.6.1. Production of green-fibers*

The dope solution consist of 7.5 $wt.\%$ Polyethersulfone (BASF Ultrason 6020 P, dried prior to use) as polymer binder, 22.5 $wt.\%$ NMP (1-Methyl-2-pyrrolidinone, 99%, extra pure, ACROS Organics$^{TM}$), and 70 $wt.\%$ titanium powder with an average particle size of 15 $\mu m$ (ASTM, Grade 2, 99.7% purity, purchased from TLS Technik GmbH & Co. (Germany)). The aqueous bore fluid was modified by adding 5 $wt.\%$ Glycerol (purity $\leq$ 98%, Ph.Eur., anhydrous., Carl Roth) as viscosity enhancer to the de-ionized water. Spinning parameters were set as listed in Table 5. Polymer solution and bore fluid flow rates have been kept constant as well as the fiber draw speed by the pulling wheel and the air gap.

|         | PES [wt. %] | Ti [wt. %] | Glycerol [wt. %] | NMP [wt. %] | Water [wt. %] |
|---------|-------------|------------|------------------|-------------|---------------|
| Polymer | 7.5         | 70         | -                | 22.5        | -             |
| Bore    | -           | -          | 5                | -           | 95            |

Table 4. Slurry composition for spinning of PES/Titanium based green-fibers

| $V_{Polymer}$ [ml/min] | $V_{Bore}$ [ml/min] | $v_{Pull}$ [mm/s] | $h_{AirGap}$ [mm] |
|------------------------|---------------------|-------------------|-------------------|
| 5.2                    | 16                  | 2.83              | 5                 |

Table 5. Spinning parameters for PES/titanium based green-fibers

*3.6.2. Thermal post-processing of green-fibers*

PES/titanium green-fibers were processed in an additional thermal treatment in order to obtain a solid porous metallic monolithic fiber. The spun PES/titanium green-fibers were sintered in a tubular furnace (Carbolite STF 16/610) under argon atmosphere. Figure 4 shows the applied temperature profile, proposed by David et al. [45]. The argon atmosphere was applied by a sweep flow of 7.5 $mL/min$, maintaining a steady supply of argon in order to not cause temperature gradients in the tubular furnace. During the first plateau the polymer binder (here PES) of the green-fiber is combusted. Reaching the second plateau the titanium powder forms sinter connections and builds a solid porous matrix. The heating rate during the heating process towards both plateaus is held constant at 5 $°C/min$. Higher heating rates cause production disruptions and lead to an inhomogeneous outcome in terms of shape conservation. Depletion of the polymer binder, and during the progression of the sintering process the fiber volume is subject to shrinkage. Depending on the temperature and sinter time, the porosity of the obtained fibers varies according to the research by David et al [45]. Here we have shown that higher sinter temperatures and longer sinter times lead to depletion of free and hidden porosity [45].

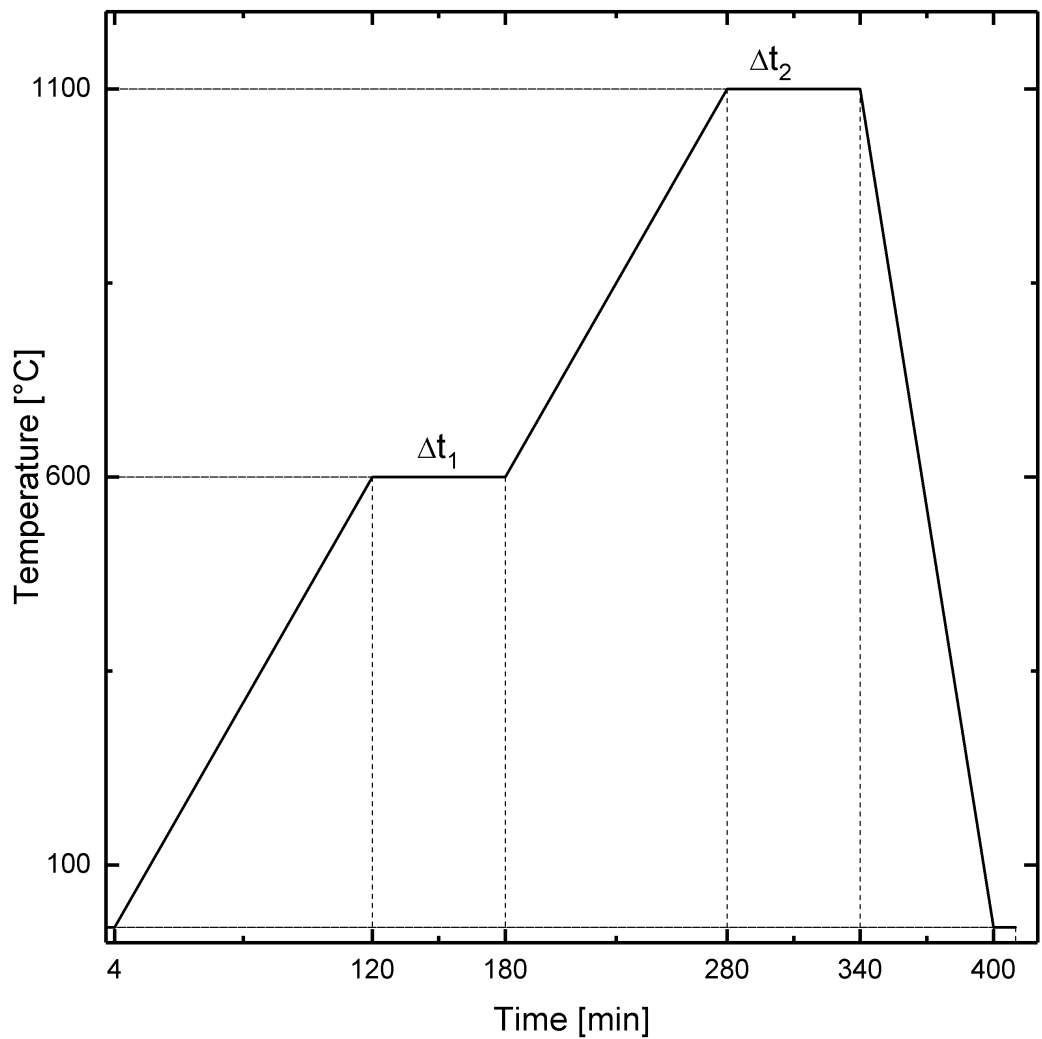

Figure 4. Temperature profile of the thermal post-processing step for PES/Titanium based green-fibers. A two plateau method is applied with defined heating rates of $5°C/min$. The temperature is held constant for 60 $min$ at 600 $°C$ for the polymer binder combustion step and at 1100 $°C$ for the sintering step.

*3.7. Optical analysis*

Photographs have been taken with a Nikon D7100 Digital Single Lens Reflex (DSLR) camera equipped with a 17-105 mm lens. Makro images of fibers (Figure 6) have been taken

T. Luelf et al. / Journal of Membrane Science 00 (2018) 1–32    16with a custom made microscope setup with a CMOS uEye$^{\text{TM}}$ LE USB 2.0 camera from iDS and a Micro Nikkor 55 mm f/2.8 Nikon lens which is mounted in retro mode.

FeSEM images have been taken with Field Immission Elektron Mikroscope Hitachi S-4800.

SEM images have been taken with Hitachi Table Top TM3030 plus. The acceleration voltage has been set to 15kV.

$\mu - CT$ images have been taken with a Bruker, SkyScan 1272 device by Bruker.

## 4. Results

### 4.1. Prototyping spinnerets with 3D-printing

As 3D-printing presents a higher degree of freedom as compared to traditional manufacturing techniques during the design process, the shape of the spinnerets was varied from the standard, concentric round, shape. In conventional spinnerets the polymer solution enters the spinneret from the top and is distributed concentrically around the needle through multiple drillings. Employing our prototype spinnerets, the bore fluid needle is connected to the outer body at one quarter, leaving three quarters of spinneret for distribution of the polymer solution. To ensure an equal distribution of polymer solution at the outlet, the polymer flow channel was narrowed directly at the opening (see Figure 5). Inner surfaces of 3D printed spinnerets comprise a higher roughness as compared to polished metal spinnerets. This is more pronounced for the spinnerets based on polyjet printing than for the stereolithography based spinnerets. On the outer surface on the other hand, this is not found for the stereolithography based parts and the glossy polyjet parts.

A comparison of designed and printed spinnerets with different outlet geometries is displayed in Figure 3. Rendering of the designed geometry is displayed for each pair on the left side in an isometric view, photographs of the printed versions are displayed on the right. No significant deviations from the intended geometry, especially at the outlet are visible. The choice of 3D-printing over conventional manufacturing techniques provides an increased freedom of design, especially concerning the spinneret outlet, where the fiber geometry is defined during the spinning process. Variations in bore fluid channel geometry and channel numbers as well as polymer solution channel geometry are possible to be obtained without increase in fabrication effort.

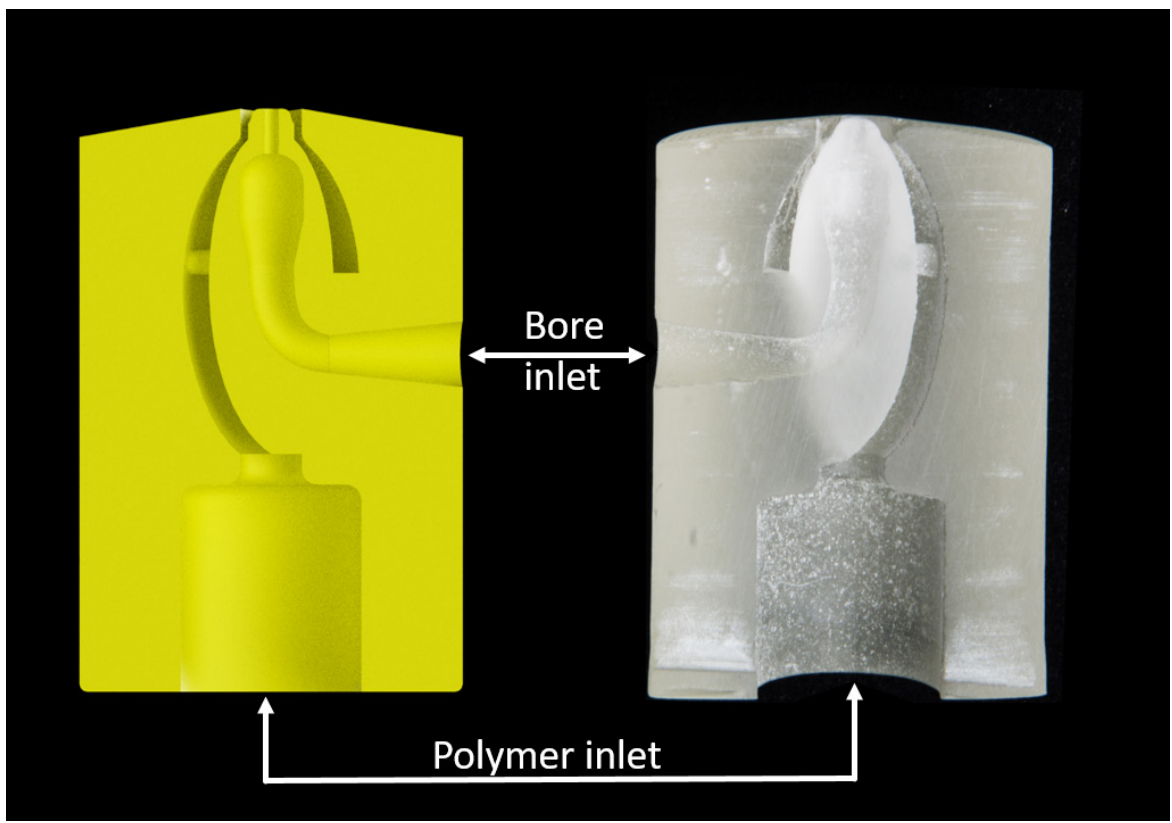

Figure 5. Close-up of the outlet region of a 3D-printed spinneret cut in half. Bore fluid and polymer solution channel maintained its structure during printing. The inner channels show a higher roughness than machined spinnerets.

*4.2. Designing tri-bore spinnerets - Influence of needle size*

The potential for iterative spinneret improvement is exemplary shown for the influence of needle size using a tri-bore spinneret. The effect was evaluated by designing two spinnerets, one with a flat bore outlet surface, but slitted polymer paths towards the center of the spinneret and the other with extruded bore needles and (Figure 6 A).

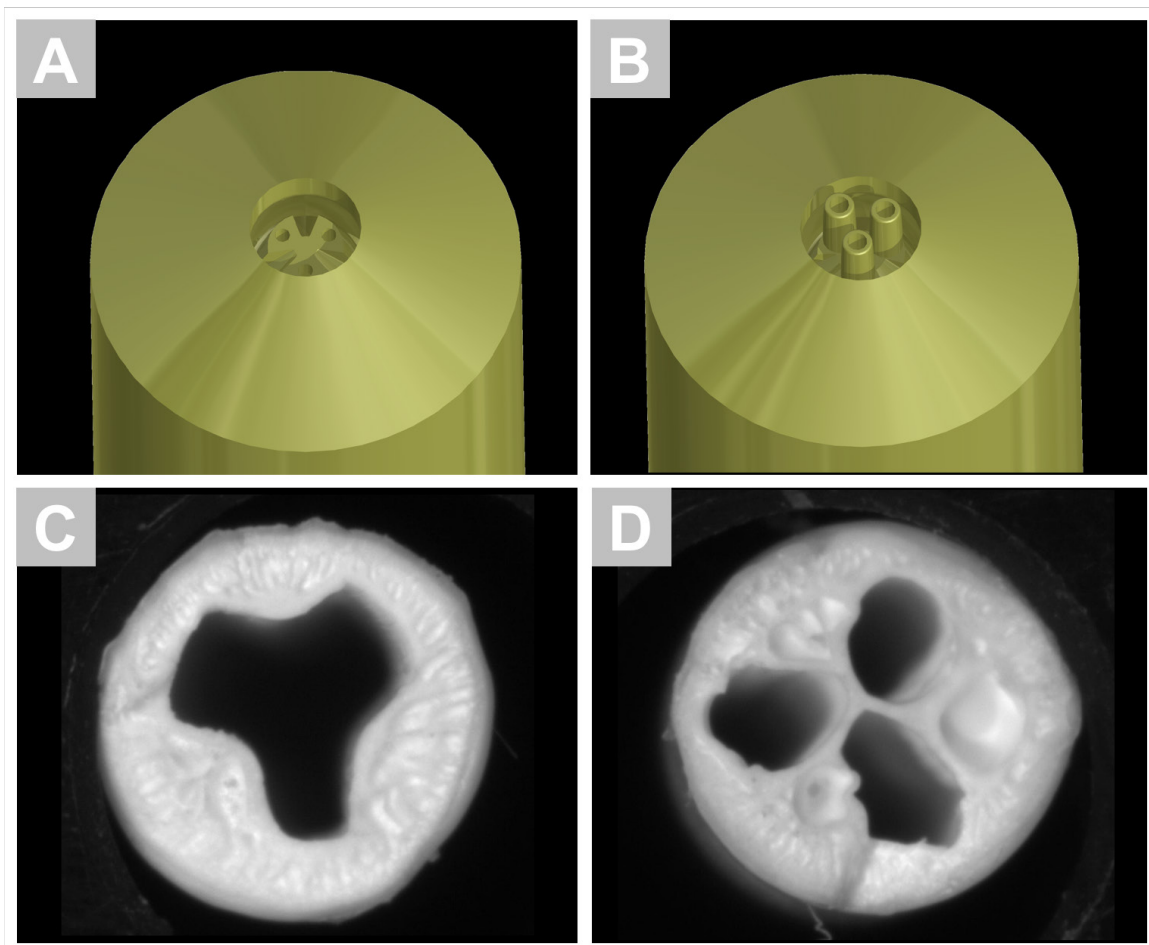

Figure 6. Influence of the needle size on fiber channel integrity. A) CAD rendering of the spinneret without extruded needles; B) CAD rendering of the spinneret with extruded needles; C) Photograph of a fiber cross-section spun with a spinneret without extruded needles (A); D) Photograph of a fiber cross-section spun with a spinneret with extruded needles.

We observed a more pronounced three-channel cross-section for the extruded needle geometry, as shown in Figure 6 D. Employing the flat bore opening the three bore channels are connected to form one channel inside the spun hollow fiber as depicted in Figure 6 C. The extruded bore needles (Figure 6 B provide a stable flow channel for the polymer dope solution in-between the bore channels of the spinneret. Therefore, this approach avoids blockage of the dope fluid's pathways to the center as occurring utilizing the flat openings (Figure 6A, C). Thus extruded bore needles tend to be more robust against interconnection of bore streams as polymer solution is guided also between the needles. In contrast, utilizing the flat bore openings there is no pathway for the polymer solution to enter the space between the bore openings again, as reported also by Wang [35] for the distance of the channels. It was found that hollow geometries at the spinneret outer surface with thin wall thicknesses of below 0.8 $mm$ are not feasible to be printed with support material that demands subsequent cleaning steps, as mechanical stress easily leads to material failure in this regions.

*4.3. Influence of rotation on tri-bore fiber morphology*

Figure 7 depicts $\mu - CT$ imaging, of the fiber structure. Figure 7 A shows the fiber material of a short sample detected by $\mu - CT$ imaging. One can clearly see the cross-section and the three channels, that are rotating around one another. In Figure 7 B and C the lumen free volume is coloured in blue to visualize the flow channels. Figure7 C is taken from a long fiber sample, revealing more information of the whole fiber with potential production instabilities. This $\mu - CT$ measurements also reveal, that certain chaotic distortions lead to radial channel interconnection (C). We account the characterization of hollow fiber membranes via $\mu - CT$ superior over FeSEM imaging in terms of integral fiber properties. Utilizing this techniques in material studies could lead to a better understanding of hollow fiber properties in the future. Figure 8 shows the cross-section of three tri-bore fibers spun with variation in rotational speed. Figure 8 Ashows a fiber, that was spun with no rotation, whereas Figure 8 B and 8 C represent fibers spun with 30 RPM and 65 RPM respectively.

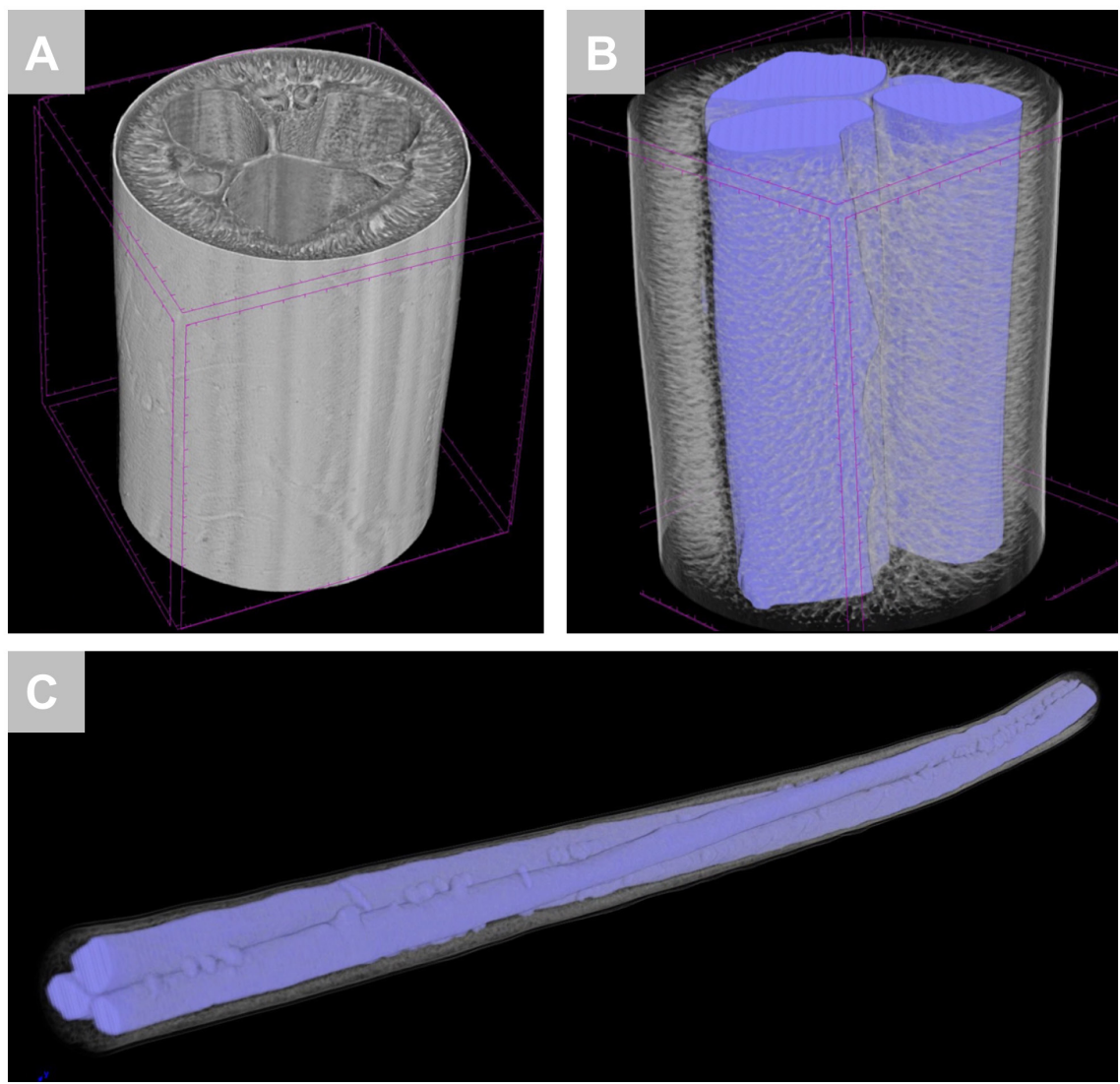

Figure 7. μ-CT imaging of rotating multibore hollow fibers spun with 30 RPM rotation of the spinneret. A) Cross-section of a short sample, B) Cross-section of a short sample with colored free volume, C) Colored free volume of a long sample.

For all samples a clear tri-bore structure is visible without channel interconnections. Macrovoids are located between the lumen channels towards the outside of the fiber. Further, finger like macrovoids are located near the fiber outer region. The inner and outer separation layers are not penetrated by the voids. Fibers spun with 30 RPM (Figure8 B) show a morphology, comparable with the fibers spun without rotation. Slight deformation of the lumen structure is visible for the sample spun with 65 RPM (Figure8 B), as one would expect for a rotating co-extrusion due to shear stress at the spinneret opening. The outside diameter of the rotating and non-rotating tri-bore fibers show no significant difference and in the final fibers rotational pitches were evaluated to be 23.3 $mm$ for the fibers spun with 65 RPM, being in good agreement with the theoretical value of 27.5 $mm$. If one applies laminar flows to these fibers with Reynolds numbers of $Re = 200$, Dean numbers of $De_{200,30RPM} = 71$ and $De_{200,65RPM} = 89$ are calculated.

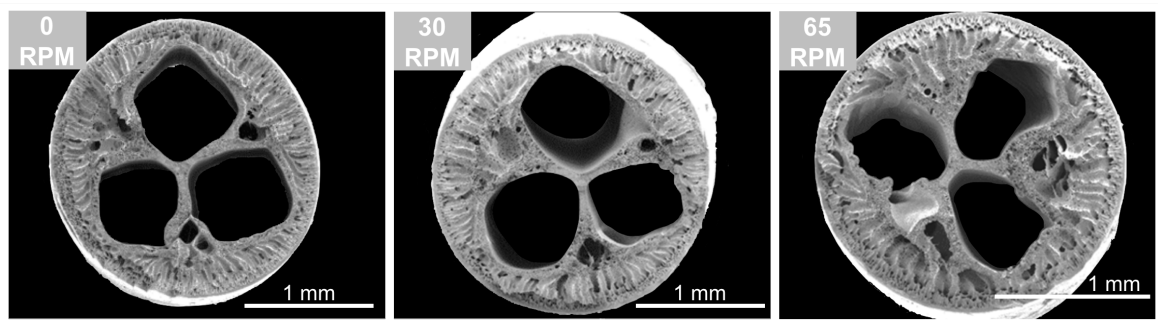

Figure 8. Cross-sectional FeSEM images of tri-bore fibers spun under rotating conditions. Fibers are spun with 0 RPM, 30 RPM and 65 RPM, respectively. The background was adapted (blackened) for better visibility of the inner fiber channels. Original FeSEM Images of the fibers can be found in the supplementary material.

*4.4. Helical tri-bore hollow fibers (PES- and titanium-based)*

The resulting helical tri-bore hollow fibers are displayed in Figure 10 and Figure 11 for classical polymer solution based PES fibers and metal-polymer slurries based fibers resulting in solid metallic monolithic membrane after thermal post-processing, respectively.

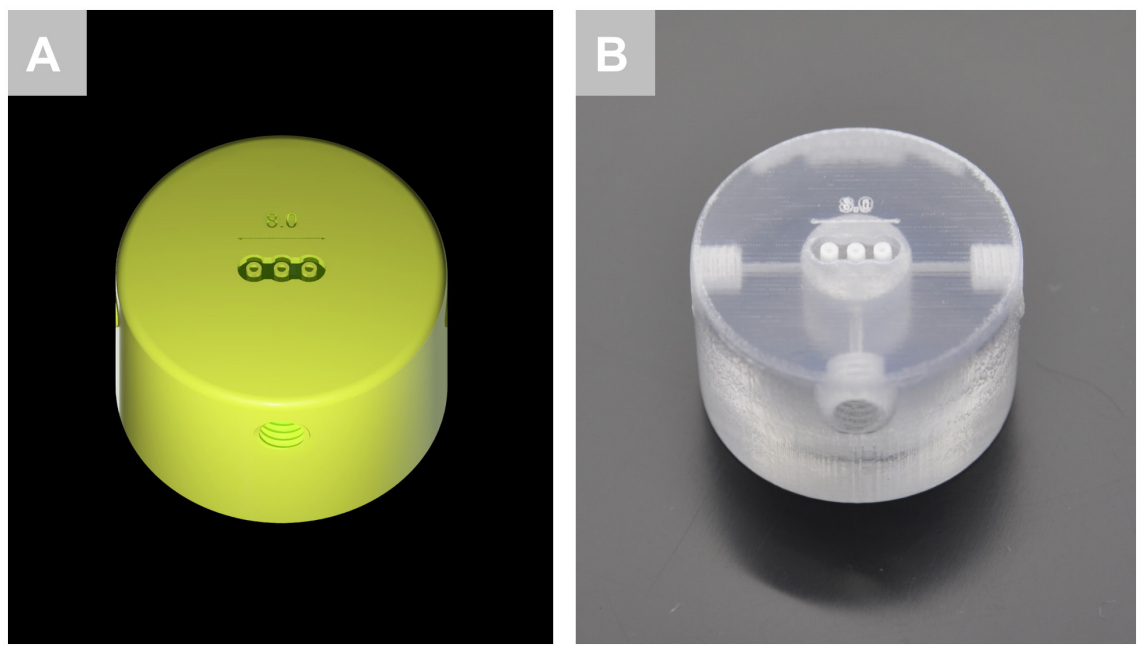

Figure 9. Multibore spinneret for helical and flat tri-bore fibers. Comparison of A) CAD rendering of the spinneret and B) Photograph taken after printing of the spinneret.

### 4.4.1. PES based helical multibore fibers

The classical polymer solution based PES multibore fibers were spun with 0 RPM to form flat tri-bore fibers and with 30 RPM to form helical tri-bore hollow fibers. Figure 10 A shows a comparison of both geometries. Figure 10 B shows the cross-section (spun with 0 RPM), presenting a similar macro-void distribution as polymeric tri-bore hollow fibers (compare Figure 8). A dense skin formation can be observed on the lumen side and the outer surface. Gradual pore size increase from the skin towards the polymer matrix is due to different solvent exchange rates. When exposed to the non-solvent, the phase inversion process takes place rapidly but becomes more and more hindered due to diffusion boundaries. The inner bore channels evolve with equal size with slight deformation. The arrangement of bore needles in the spinneret promote the formation of stabilizing grooves on the outer side of the fiber. This effect is enhanced when rotation is applied. The formed grooves show resemblance to buckling effects of polymer layers on soft substrate [48]. The rotational pitches were evaluated to be 6.50 *mm* with the spinning parameters applied (Table 3). Figure 10 C shows a uniform pitch along the whole fiber. The theoretical pitch was calculated to $pitch_{th} = 6.90$ *mm* and shows good agreement with the actual value.

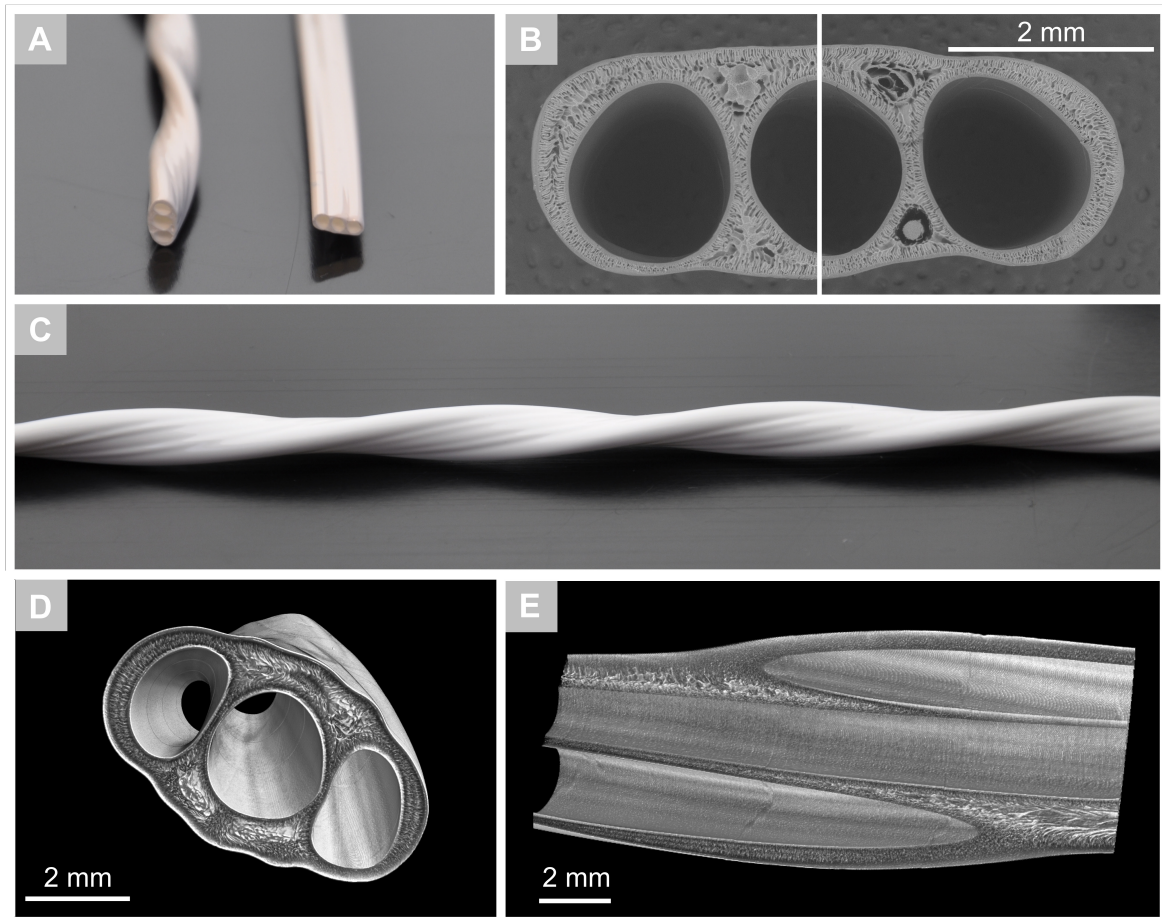

Figure 10. Helical tri-bore fibers PES based: A) Comparison of sample spun with 30 RPM (left) and sample spun with 0 RPM (right) B) Cross-section of sample spun with 0 RPM. The picture is composed of two seperate images with the same magnification to capture the whole cross-section. C) Side view of the sample spun with 30 RPM showing a uniform pitch along the whole fiber length.

*4.4.2. Metal based helical multibore fibers*

The metal-polymer slurries based fibers resulting in solid metallic monolithic membranes were spun with 0 RPM to form flat tri-bore fibers and with 30 RPM to form helical tri-bore hollow fibers. Figure 11 A shows a comparison of both geometries in the green form and after thermal post processing. The stable helical and flat geometry maintained its shape after extrusion and after consecutive thermal post-processing. The green-fiber is subjected to shrinkage during the thermal post-processing step. Shrinkage occurs due to combustion of the binding polymer of the green-fiber as well as the progressing sinter stage. The difference in structure can be observed in the SEM image of Figure 11 B showing the cross-section of the green-fiber (bottom) and fiber after thermal post-processing (top). Corresponding larger magnification depicted in Figure 11 C shows the titanium particles embedded between the PES polymer binder and in Figure 11 D the fiber after thermal post-processing with clearly visible sintered particles without polymer binder which form a solid porous titanium matrix. The rotational pitches of the fibers were evaluated to be 6.50 $mm$ for the green-fiber with the spinning parameters applied (Table 5). A uniform pitch along the whole fiber can be observed after extrusion as shown in Figure 11 A. The form of the pitch is maintained after the thermal post-processing step, and reduced due to shrinkage to 6.30 $mm$. The theoretical pitch was calculated to $pitch_{th} = 5.65$ $mm$. Comparing the spinning parameters and the pitch developed, we experienced that the spinning system is very sensitive towards small changes in parameters and material systems. The curved geometry allow for the calculation of Dean numbers. If one applies flow inside the channels with Reynolds numbers of $Re = 200$, the Dean numbers were calculated to $De_{200,30RPM} = 105$ and $De_{200,30RPM} = 199.9$ for the PES based and sintered fibers respectively.

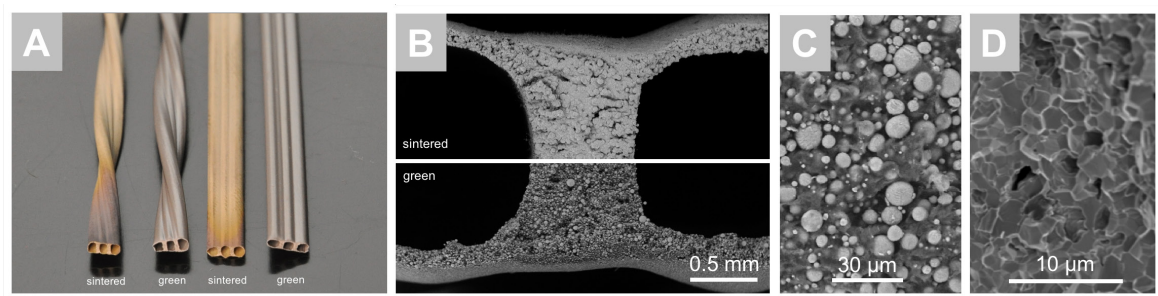

Figure 11. Helical tri-bore fibers titanium-based: A) Comparison of green-fibers and fibers after thermal post-processing spun at 0 RPM and 30 RPM, respectively. B) SEM image of the cross-section of the green-fiber (bottom) and fiber after thermal post-processing (top) C) Magnified image of the green-fiber showing the Titanium particles with polymer binder D) Magnified image of the fiber after thermal post-processing with clearly visible sintered particles without polymer binder.

## 4.5. Channel integrity

Figure 12 shows the influences of spinning process parameters on the key geometric properties of the helical multibore fiber. Typical machine parameters, such as volume flows, air gap distance and pulling speed act on the channel dimensions as in hollow fiber production with simple bore. The key differences are to be found in the channel integrity and being influenced by multiple parameters. In this regard a core aspect figured during the conduction of the here presented experiments is the interplay of phase separation, viscosities, and spinneret design. All aspects are influencing the channel integrity, which is the stability of fiber design during fabrication. If, for instance, the flow of bore fluid is not constant in all channels of the spinneret, bore channels with different diameters evolve. Additionally, multiple bore channels tend to combine to one channel if the geometry is not stabilized. The stability can be generated by two means. First, fast phase separation already takes place in the air-gap region, which is also coupled to the risk of blocking of the spinneret causing fiber rupture. Secondly, an enhancement of the bore fluid's viscosity hinders mobility of the polymer solution - bore liquid interfaces. Furthermore, the spinneret design is a key aspect to stabilize multiple channel geometries, as explained in Section 4.2.

Besides the amount of round channels, the spinneret design also enables the formation of structured channels, as investigated by Çulfaz [24]. Nevertheless, our approach offers a more flexible route to geometry development compared to the form-defining spinneret parts being produced employing laser ablation. A combination with the rotational fiber production approach would finally offer a way to not only increase specific surface area of lumen channels but also go beyond an alignment that is parallel to the main direction of flow, thus offering flow disturbance.

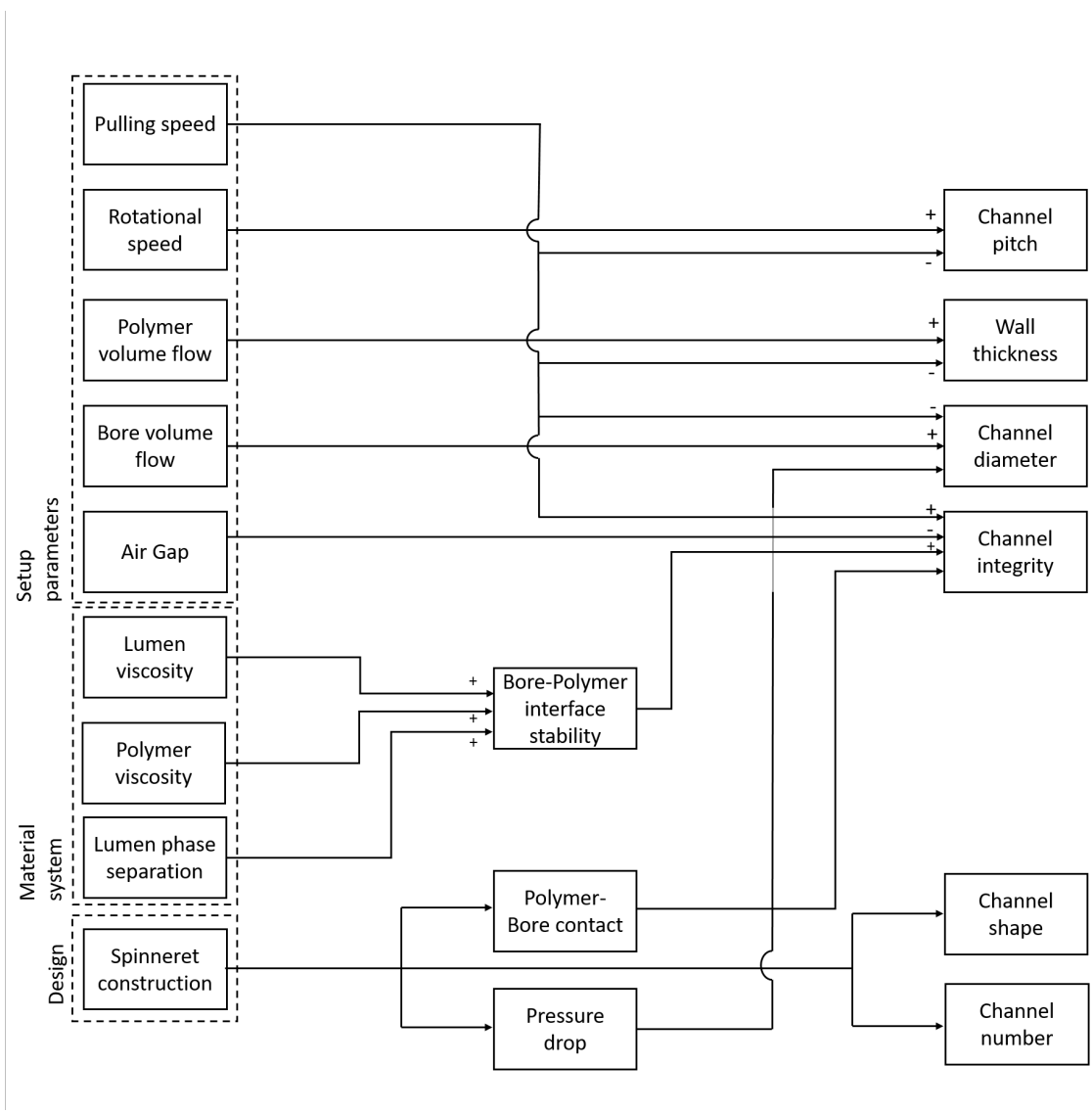

Figure 12. Spinning related influences on the fiber structure during rotating spinning.

## 5. Conclusion

In this manuscript we present a hands-on toolbox for spinneret and hollow fiber membrane development. We show 3D printing applications in spinneret design and construction for different available printing materials as well as a validation of stability against NMP as a typical solvent in phase inversion membrane formation. Examples of iterative spinneret optimization with a variation of fiber geometry are presented for polymeric fibers. The produced multibore fibers have a stable 3D geometry with separated lumen channels. In addition we take the concept of spinneret design to another level by combining the typically implemented co-extrusion with the design of a rotation assembly. The resulting fibers combine a multibore approach with high packing density with a fiber lumen rotation enabling dean vortice formation during operation. Subsequently, we applied the rotating spinning method to titanium loaded polymer solutions. The phase separated green-fibers preserve the spiralling multibore structure during thermal post processing. This enables the development and production of complex 3D structured metal fibers for electrochemical membrane processes. With this approach we are able to fabricate fibers with rotating inner or outer geometries. Thus, on the basis of our research one is able to tailor both the membrane inner and outer mixing properties to a separation task.

**Acknowledgement**

M.W. acknowledges the support through an Alexander-von-Humboldt Professorship. This work was performed in part at the Center for Chemical Polymer Technology CPT, which is supported by the EU and the federal state of North Rhine-Westphalia (grant no. EFRE 30 00 883 02). This project has received funding from the European Research Council (ERC) under the European Unions Horizon 2020 research and innovation program (grant agreement no. 694946) and German Federal Ministry of Education and Research (BMBF) under the project $Tubulair\pm$ (03SF0436B). The authors would like to thank Karin Faensen for her help with sample preparation and electron microscopy.

*T. Luelf et al. / Journal of Membrane Science 00 (2018) 1–32* 30[17] P. Moulin, J. Rouch, C. Serra, M. Clifton, P. Aptel, Mass transfer improvement by secondary flows: Dean vortices in coiled tubular membranes, Journal of Membrane Science 114 (2) (1996) 235 – 244.

[18] S. Luque, H. Mallubhotla, G. Gehlert, R. Kuriyel, S. Dzengeleski, S. Pearl, G. Belfort, A new coiled hollow-fiber module design for enhanced microfiltration performance in biotechnology, Biotechnology and Bioengineering 65 (3) (1999) 247–257.

[19] R. Moll, D. Veyret, F. Charbit, P. Moulin, Dean vortices applied to membrane process: Part I. Experimental approach, Journal of Membrane Science 288 (1) (2007) 307 – 320.

[20] R. Moll, D. Veyret, F. Charbit, P. Moulin, Dean vortices applied to membrane process: Part II: Numerical approach, Journal of Membrane Science 288 (1) (2007) 321 – 335.

[21] P. Manno, P. Moulin, J. Rouch, M. Clifton, P. Aptel, Mass transfer improvement in helically wound hollow fibre ultrafiltration modules: Yeast suspensions, Separation and Purification Technology 14 (1) (1998) 175 – 182.

[22] S. Armbruster, O. Cheong, J. Lölsberg, S. Popovic, S. Yüce, M. Wessling, Fouling Mitigation in Tubular Membranes by 3D-printed Twisted Tape Turbulence Promoters, Submitted to Journal of Membrane Science.

[23] P. Z. Çulfaz, E. Rolevink, C. van Rijn, R. G. H. Lammertink, M. Wessling, Microstructured hollow fibers for ultrafiltration, Journal of Membrane Science 347 (1) (2010) 32 – 41.

[24] P. Z. Çulfaz, M. Wessling, R. G. H. Lammertink, Hollow fiber ultrafiltration membranes with microstructured inner skin, Journal of Membrane Science 369 (1) (2011) 221 – 227.

[25] P. Z. Çulfaz, M. Wessling, R. G. H. Lammertink, Fouling behavior of microstructured hollow fiber membranes in submerged and aerated filtrations, Water Research 45 (4) (2011) 1865 – 1871.

[26] T. Luelf, C. Bremer, M. Wessling, Rope coiling spinning of curled and meandering hollow-fiber membranes, Journal of Membrane Science 506 (2016) 86–94.

[27] T. Luelf, M. Tepper, H. Breisig, M. Wessling, Sinusoidal shaped hollow fibers for enhanced mass transfer, Journal of Membrane Science 533 (2017) 302–308.

[28] P. Moulin, P. Manno, J. Rouch, C. Serra, M. Clifton, P. Aptel, Flux improvement by Dean vortices: ultrafiltration of colloidal suspensions and macromolecular solutions, Journal of Membrane Science 156 (1) (1999) 109 – 130.

[29] M. Wiese, O. Nir, D. Wypysek, M. Wessling, Fouling minimization at membranes having a 3D surface topology with microgels as soft model colloids, Submitted to Journal of Membrane Science.

[30] J. de Jong, N. Benes, G. Koops, M. Wessling, Towards single step production of multi-layer inorganic hollow fibers, Journal of Membrane Science 239 (2) (2004) 265 – 269.

[31] M. S. Tijink, M. Wester, J. Sun, A. Saris, L. A. Bolhuis-Versteeg, S. Saiful, J. A. Joles, Z. Borneman, M. Wessling, D. F. Stamatialis, A novel approach for blood purification: Mixed-matrix membranes30